\documentclass[a4paper,8pt]{article}

\usepackage[english]{babel}

\usepackage{natbib}

\usepackage{amsfonts,amsbsy,amssymb,amsmath,graphicx,float} 

\usepackage[paperwidth=18cm,paperheight=29.7cm,hmargin=2cm,vmargin=3cm]{geometry}

\usepackage[linktoc=all,colorlinks=true,citecolor=blue]{hyperref}

\begin{document}
	
	\title{An Extension of Discrete Lagrangian Descriptors \\ for Unbounded Maps}
	
	\author{V\'{i}ctor J. Garc\'{i}a-Garrido$^{1}$ \\[.3cm]
	$^1$ Departamento de F\'{i}sica y Matem\'{a}ticas, \\ Universidad de Alcal\'{a}, 28871, Alcal\'{a} de Henares, Spain. \\[.2cm] vjose.garcia@uah.es}

\maketitle

\begin{abstract}

In this paper we provide an extension for the method of Discrete Lagrangian Descriptors with the purpose of exploring the phase space of unbounded maps. The key idea is to construct a working definition, that builds on the original approach introduced in \citet{carlos}, and which relies on stopping the iteration of initial conditions when their orbits leave a certain region in the plane. This criterion is partly inspired by the classical analysis used in Dynamical Systems Theory to study the dynamics of maps by means of escape time plots. We illustrate the capability of this technique to reveal the geometrical template of stable and unstable invariant manifolds in phase space, and also the intricate structure of chaotic sets and strange attractors, by applying it to unveil the phase space of a well-known discrete time system, the H\'{e}non map.

\end{abstract}

% \pacs{needs pacs number}

\maketitle


% \tableofcontents

\section{Introduction}
\label{sec:intro}

The method of Lagrangian Descriptors (LDs) is a trajectory-based scalar diagnostic originally developed in the context of Geophysics \cite{mendoza2010} to study Lagrangian transport and mixing processes that take place in the ocean or the atmosphere. This tool was first introduced a decade ago in the work by \citet{madrid2009} as a tool to detect \textit{Distinguished Hyperbolic Trajectories} (or 'moving saddles') in dynamical systems with general time dependence, and in recent years, it has been found to be an excellent technique for exploring and revealing the geometrical template of invariant manifolds \cite{mancho2013lagrangian,lopesino2017,naik2019a} that characterizes phase space regions with qualitatively distinct dyamical behavior, a long-sought goal envisioned by Henri Poincar\'{e} in his works on the three-body problem in Celestial Mechanics \cite{hp1890}.

Lagrangian Descriptors have been widely applied in the literature to problems that arise in many scientific disciplines. For instance, it has been used in Oceanography for the assessment of marine oil spills \citep{gg2016}, and also to plan autonomous underwater vehicle transoceanic missions \cite{ramos2018}. Another field where LDs have received recognition is in Chemistry \cite{craven2015lagrangian,JH2015}. In Transition State Theory, which deals with the study of chemical reaction dynamics, the knowledge of phase space structures is crucial to analyze reaction rates. In this framework, the tool has proved to successfully identify normally hyperbolic invariant manifolds of Hamiltonian systems \cite{demian2017,naik2019a,naik2019b}, which serve as a scaffolding for the construction of the dividing surface to measure reation rates \cite{craven2016deconstructing, craven2015lagrangian, craven2017lagrangian, junginger2017chemical,feldmaier2017obtaining}. In all these contexts, the dynamical system that governs the long-term evolution of particles is continuous in time, and the vector field that defines it is given either by an analytical expression or by a dataset generated from a numerical model. 

In the discrete-time setting, LDs have also been defined for maps \citep{carlos}, where a mathematical connection was established between the invariant geometrical structures present in a map's phase space and the output obtained from the method. In fact, Discrete Lagrangian Descriptors (DLDs) have provided insightful results when used to describe 2D chaotic maps such as the H\'{e}non, Lozi, and Arnold's cat maps \citep{carlos,carlos2,GG2018RCD2}. In all these examples which display a strong mixing in a given area of their domain, DLDs are capable of uncovering the chaotic saddles responsible for the chaotic dynamics. 

However, when Lagrangian Descriptors (both in their continous- or discrete-time versions) are used to explore the phase space structures of unbounded dynamical systems, issues might arise when trajectories of initial conditions escape to infinity in finite time or at an increasing rate. This behavior will result in NaN values in the LD scalar field, obscuring the detection of invariant manifolds. Recent studies \cite{junginger2017chemical,naik2019b,GG2019} have revealed this concern for some examples of unbounded continuous-time dynamical systems, where the approach of calculating LDs by integrating initial conditions on a phase space grid for the same time, known as fixed-time LDs, does not work well because of escaping trajectories. In order to circumvent this issue, an alternative definition of LDs was introduced for continuous-time systems, where the value of LDs for any initial condition is calculated by evolving its trajectory for a chosen integration time or until it leaves a sufficiently large domain in phase space. The goal of this paper is to adapt this definition for discrete-time systems, and demonstrate that by considering variable iteration number Discrete Lagrangian Descriptors (VIN-DLD) one can analyze in full detail the phase space structures that govern the dynamics of two-dimensional unbounded maps.  

This paper is organized as follows. In Section \ref{sec:sec1} we briefly describe the original definition of the method of Discrete Lagrangian Descriptors (DLD) and propose a modified version of it to make it work for unbounded maps. The idea behind this alternative definition is to stop the iteration of the orbits of initial conditions that leave a certain fixed phase space region. This implies, in particular, that the orbits of initial conditions might be iterated for a different number of iterations. Section \ref{sec:res} is devoted to discussing the results of this work. First, we numerically show how the VIN-DLD succeeds in revealing the geometrical template of invariant manifolds in the phase space for the H\'{e}non map. Moreover, we illustrate that this technique is also capable of detecting KAM tori, and, as a validation, we compare this diagnostic with the classical average-exit time distribution approach \cite{meiss1997}. To finish, we demonstrate that VIN-DLD have the capability of unveiling the strange attractor that appears in the non area-preserving H\'{e}non map. Finally, in Section \ref{sec:conc} we present the conclusions of this work.

\section{Variable Iteration Number Discrete Lagrangian Descriptors}
\label{sec:sec1}

We begin this section by briefly explaining the basic setup for the method of Discrete Lagrangian Descriptors, that was first introduced in \citet{carlos} for 2D area-preserving maps. Consider a domain $D \subset \mathbb{R}^{2}$ and a function $f$ that defines the discrete-time dynamical system,
\begin{equation}
	\mathbf{x}_{i+1} = f(\mathbf{x}_{i}) \quad,\quad \text{with} \quad \textbf{x}_i = (x_i,y_i) \in D \quad,\quad \forall i \in \mathbb{N} \cup \lbrace0\rbrace
	\;.
	\label{discrete_DS}
\end{equation}
We assume that this mapping is invertible so that,
\begin{equation}
	\mathbf{x}_{i} = f^{-1}(\mathbf{x}_{i+1}) \quad,\quad \text{with} \quad \textbf{x}_i = (x_i,y_i) \in D \quad,\quad \forall i \in \mathbb{Z}^{-}
	\label{discrete_DS_back}
\end{equation}

Given a fixed number of iterations $N > 0$ of the maps $f$ and $f^{-1}$, and take $p \in (0,1]$ to specify the $L^p$-norm that we will use, then the DLD applied to systems (\ref{discrete_DS}) and (\ref{discrete_DS_back}) is given by the following function,
\begin{equation}
	MD_{p}\left(\mathbf{x}_0,N\right) = \sum_{i=-N}^{N-1} ||\textbf{x}_{i+1}-\textbf{x}_{i}||_{p} = \sum_{i=-N}^{N-1} |x_{i+1}-x_{i}|^{p} + |y_{i+1}-y_{i}|^{p}
	\label{DLD}
\end{equation}
where $\mathbf{x}_0 \in D$ is any initial condition. Observe that Eq. (\ref{DLD}) can be split into two quantities,
\begin{equation}
MD_{p}\left(\mathbf{x}_0,N\right) = MD_{p}^{+}\left(\mathbf{x}_0,N\right) + MD_{p}^{-}\left(\mathbf{x}_0,N\right) \;, 
\end{equation}
where,
\begin{equation}
	MD_{p}^{+} = \sum_{i=0}^{N-1} ||\textbf{x}_{i+1}-\textbf{x}_{i}||_{p} = \sum_{i=0}^{N-1} |x_{i+1}-x_{i}|^{p} + |y_{i+1}-y_{i}|^{p},
	\label{DLD+}
\end{equation}
measures the forward evolution of the orbit of $\mathbf{x}_0$ and,
\begin{equation}
	MD_{p}^{-} = \sum_{i=-N}^{-1} ||\textbf{x}_{i+1}-\textbf{x}_{i}||_{p} = \sum_{i=-N}^{-1} |x_{i+1}-x_{i}|^{p} + |y_{i+1}-y_{i}|^{p},
	\label{DLD-}
\end{equation}
accounts for the backward evolution of the orbit of $\mathbf{x}_0$. In this framework, \cite{carlos} provides a mathematical proof that DLDs highlight stable and unstable manifolds at points where the $MD_p$ field becomes non-differentiable (discontinuities or unboundedness of its gradient). These non-differentiable points are displayed in a simple way when depicting the $MD_p$ scalar field, and are known in the literature as ''singular features''. Therefore, DLDs are capable of recovering the geometrical template of invariant stable and unstable manifolds present in the phase space of any map defined by Eqs. (\ref{discrete_DS}) and (\ref{discrete_DS_back}). Moreover, \cite{carlos} proves for several examples that $MD_{p}^{+}$ detects the stable manifolds of the fixed points and $MD_{p}^{-}$ does the same for the unstable manifolds of the fixed points. Therefore, the fixed points with respect to the map (\ref{discrete_DS}) will be located at the intersections of these structures.

In the definition of DLDs, the number of iterations $N$ considered for the computation of the orbit of any initial condition plays a crucial role for revealing the intricate structure of stable and unstable manifolds in the phase space, and hence holds the \textit{secret} to unlock the general applicability of DLDs to unbounded maps. It has been shown in \cite{carlos} that, as the number of iterations $N$ is increased, a richer template of geometrical phase space structures is recovered by the method, since we are including into the analysis more information about the past and future history of the orbits of the map. The issue arises when we deal with an unbounded map for which orbits can escape to infinity at an increasing rate. This happens for instance in the H\'{e}non map, which is the example we will use to illustrate how the definition of DLDs has to be modified in order to make the method work for general open maps. Notice that in the original definition of DLDs given in Eq. \eqref{DLD}, all the initial conditions chosen in the domain $D \subset \mathbb{R}^{2}$ are iterated using $f$ and also $f^{-1}$ for the same number of iterations $N$. Consequently, for an unbounded map, some initial conditions might escape to infinity very fast, yielding extremely large and $NaN$ values in the computation of $MD_p$, and this issue obscures completely the detection of invariant manifolds in some regions of phase space.  

The modification that we propose for the definition of DLDs in order to avoid the problem of orbits escaping to infinity in unbounded maps, which would end up rendering the $MD_p$ scalar field values useless, is the following. Start by fixing a planar region $\mathcal{R} \in \mathbb{R}^2$ in the phase space, that we call the \textit{interaction region}, and take a fixed number of iterations $N > 0$. Given an initial condition $\mathbf{x}_0 \in D$, we calculate $MD_p$ along its orbit as we iterate it with $f$ and $f^{-1}$, and we do so until we reach the maximum number of iterations $N$ or the orbit leaves the interaction region $\mathcal{R}$. Therefore, we define the numbers,
\begin{equation}
N^{\pm}_{\mathbf{x}_0} = \max_{k = 1,\ldots,N} \; \lbrace k \;|\; f^{\pm k}\left(\mathbf{x}_0\right) \in \mathcal{R} \rbrace
\label{var_its}
\end{equation}
where $N^{+}_{\mathbf{x}_0}$ and $N^{-}_{\mathbf{x}_0}$ are, respectively, the number of forward ($f$) and backward ($f^{-1}$) iterations of the orbit $\mathcal{O}\left(\mathbf{x}_0\right)$. As a result, initial conditions might be iterated a different number of times depending on whether they remain inside the interaction domain $\mathcal{R}$ or they eventually escape it. We use this idea to define what we call Variable Iteration Number Discrete Lagrangian Descriptor (VIN-DLD). This modified version of DLDs is given by the formula,
\begin{equation}
\mathcal{D}_{p}\left(\mathbf{x}_0,N\right) = \sum_{i=-N^{-}_{\mathbf{x}_0}}^{N^{+}_{\mathbf{x}_0}-1} ||\textbf{x}_{i+1}-\textbf{x}_{i}||_{p} = \sum_{i=-N^{-}_{\mathbf{x}_0}}^{N^{+}_{\mathbf{x}_0}-1} |x_{i+1}-x_{i}|^{p} + |y_{i+1}-y_{i}|^{p} \; .
\label{VT_DLD}
\end{equation}
Notice that if one takes a large interaction region $\mathcal{R}$ of the planar phase space, the expected result is that this new variable iteration number definition of DLD approaches the values of the original DLD in Eq. \eqref{DLD}. Therefore, the same mathematical properties that were proved for the DLD in \cite{carlos} also hold in this case, that is, the modified VIN-DLD also has the capability of detecting the stable and unstable manifolds of fixed points and periodic orbits at locations where $\mathcal{D}_{p}$ is non-differentiable. 

To finish, we would also like to remark that, since the interaction region $\mathcal{R}$ has to be chosen large enough so that the values of the VIN-DLD approach those obtained with the Fixed Iteration Number Discrete Lagrangian Descriptor (FIN-DLD) in Eq. \eqref{DLD}, the shape of $\mathcal{R}$ is not important in order for the method to work. For the purpose of simplifiyng the analysis, and without loss of generality, it is convenient to choose $\mathcal{R}$ as a circle centered about the origin, but a square would work perfectly fine too. In particular, to analyze the phase space structures present in the H\'{e}non that we discuss next we will use a circle of radius $r = 100$.

\section{Results}
\label{sec:res}

In this section we present the results of applying the VIN-DLD defined in Eq. \eqref{VT_DLD} to analyze the phase space structures of a well-known example with unbounded behavior, the H\'{e}non map \cite{henon1976}, which is a hallmark for studying dynamical properties of two-dimensional discrete-time systems. This invertible map is given by the system,
\begin{equation}
\begin{cases}
x_{n+1} = A + B y_n - x_{n}^2  \\
y_{n+1} = x_n
\end{cases} , \quad \forall n \in \mathbb{N} \cup \lbrace 0 \rbrace
\label{henon_map}
\end{equation}
where $A,B \in \mathbb{R}$ are the model parameters, and the inverse mapping is:
\begin{equation}
\begin{cases}
x_{n} = y_{n+1} \\[.2cm]
y_{n} = \dfrac{x_{n+1} - A + y_{n+1}^2}{B}
\end{cases} , \quad \forall n \in \mathbb{Z}^{-} \cup \lbrace 0 \rbrace
\label{henon_map_inv}
\end{equation}
The H\'{e}non map is an area preserving map when $|B| = 1$, and is orientation-preserving for $B = -1$. Moreover, in \cite{devaney1979} it is shown that, if $A > \left(5+2\sqrt{5}\right)\left(1+|B|\right)^2 / 4$, then the mapping has a hyperbolic invariant Cantor set which is topologically conjugate to a Bernoulli shift on two symbols, i.e. it has a chaotic saddle. In order to illustrate that the viariable-iteration DLD recovers all the relevant phase space invariant manifolds, we will take a look at three pairs of values for the model parameters of the H\'{e}non map. First, we will consider the case $A = 9.5$ and $B = -1$ for which the map has a chaotic saddle. Second, we discuss the dynamical structures for $A = 0.298$ and $B = 1$, and for this case we show how DLDs detect KAM tori and also the stable and unstable manifolds. Finally, we demosntrate that the method succeeds in revealing the intricate strucutre of the strange attractor that appears in the system for the classical value of the parameters $A = 1.4$ and $B = 0.3$, which was first discovered by H\'{e}non in \cite{henon1976}. 

%%%%%%%%%%%%%%%%%%%%%%%%%%%%%%%%%%%%%%%%%%%%%%%%%%%%%%%%%%%%%%%%%%%%%%%%%%%%%
%
% The most general Henon map is
%
% x' = A + B y_n - C x_{n}^2
% y' = D x_n
%
% The case A = B = 1, C = C0, D = D0 is topologically equivalent 
% to A = C0, B = D0, C = 1, D = 1 so that the Henon map:
%
% x' = A + By - x^2
% y' = x
%
% is topologically equivalent to 
%
% x' = 1 + y - Ax^2
% y' = Bx
%
% see Devaney (1979)
%
%%%%%%%%%%%%%%%%%%%%%%%%%%%%%%%%%%%%%%%%%%%%%%%%%%%%%%%%%%%%%%%%%%%%%%%%%%%%%

\begin{figure}[htbp]
	\begin{center}
		A)\includegraphics[scale=0.3]{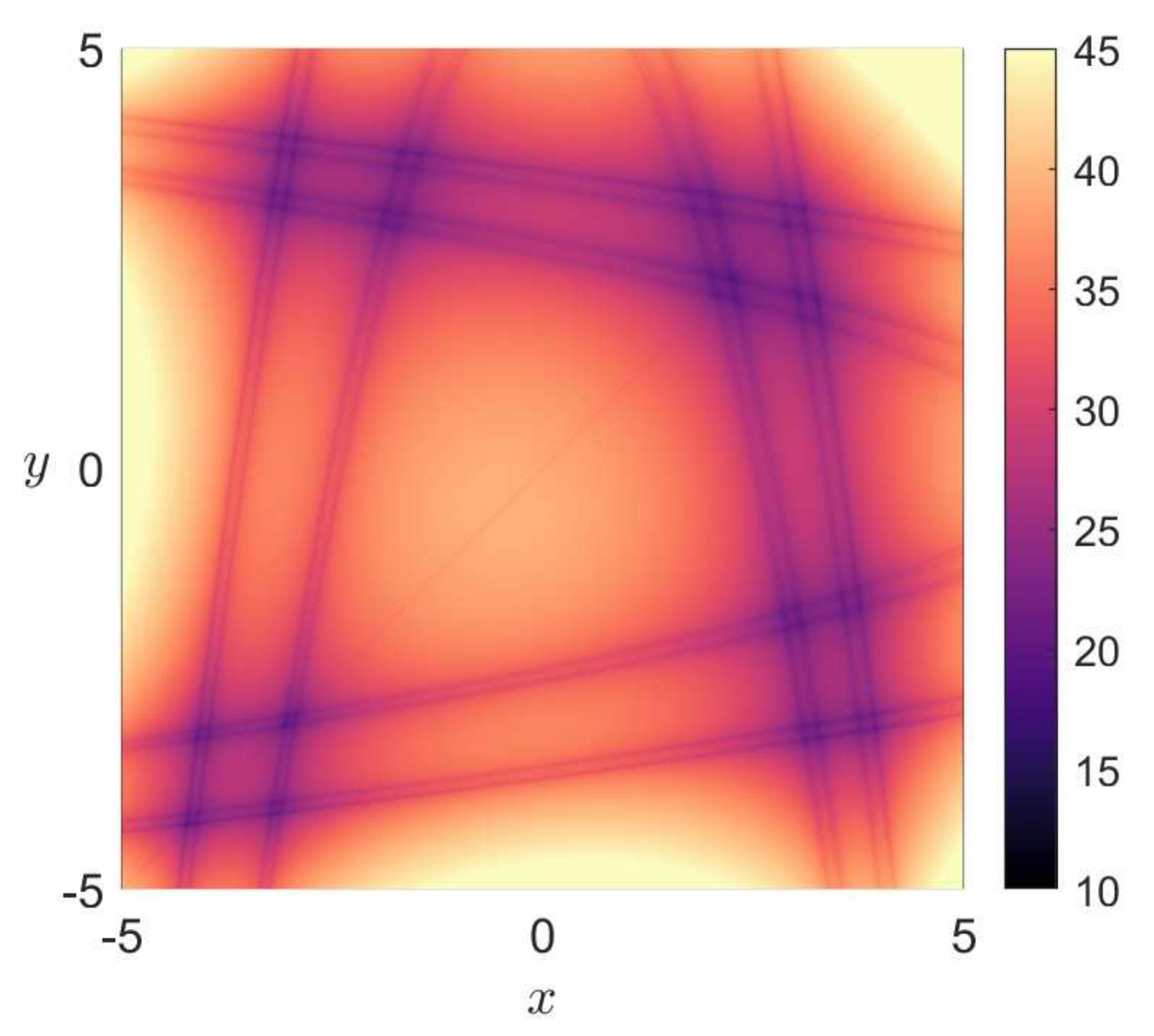}
		B)\includegraphics[scale=0.3]{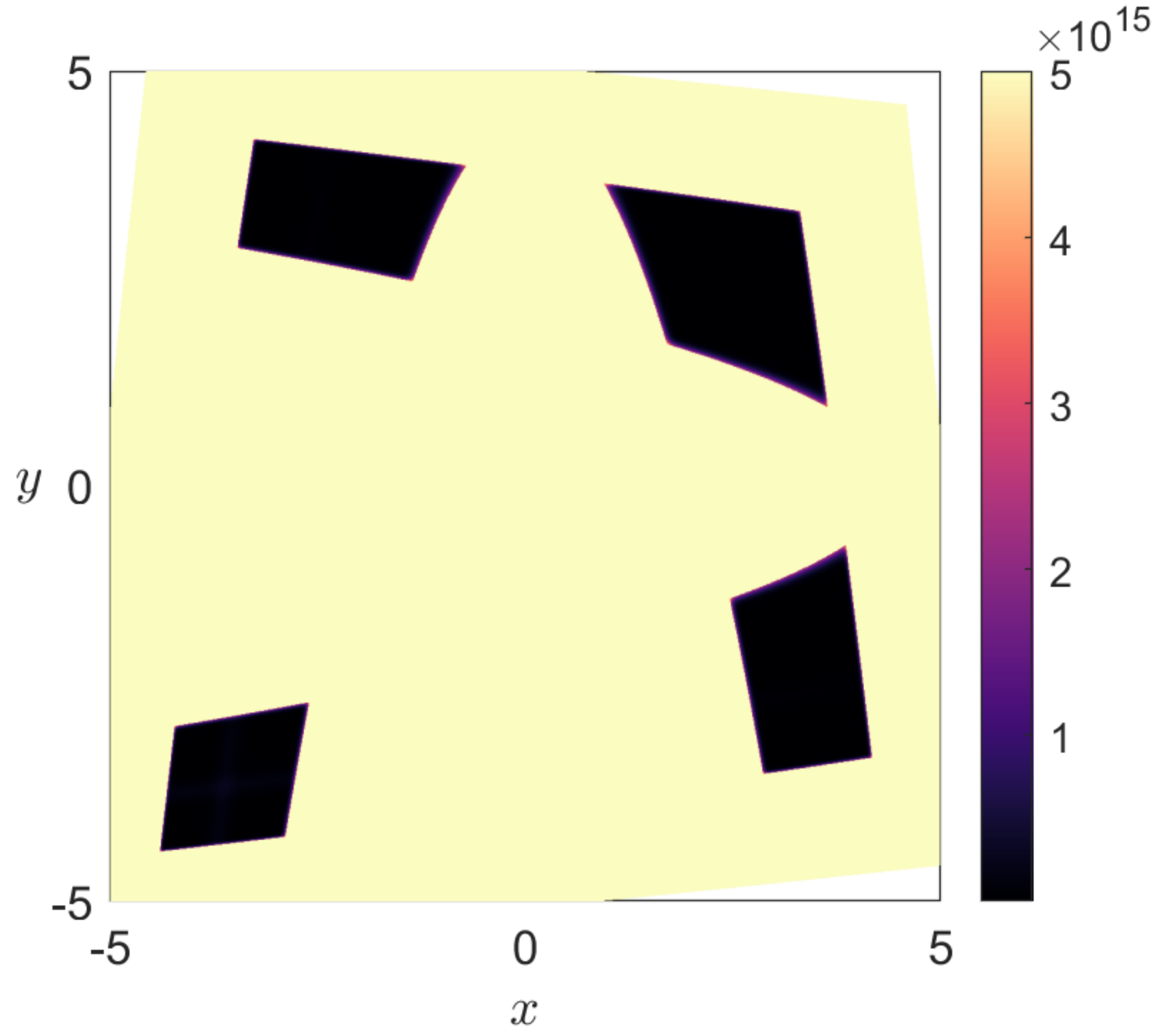}
	\end{center}
	\caption{Phase space structures obtained for the H\'{e}non map with model parameters $A = 9.5$ and $B = -1$ using the FIN-DLD with $p = 0.05$. A) $N = 5$ iterations. B) $N = 10$ iterations. For $N = 5$ the method captures the basic features of the chaotic saddle. However, for $N = 10$ iterations many orbits have already escaped to infinity, giving extremely large $MD_p$ values and rendering the diagnostic useless in this case.}
	\label{DLD_fixIt}
\end{figure}

We begin our analysis by applying the FIN-DLD introduced in Eq.\eqref{DLD} to the map with parameters $A = 9.5$ and $B = -1$. This computation was already carried out in \cite{carlos}, but the $p$-norm chosen to reveal the chaotic saddle was $p = 0.05$. The reason for choosing such a small value for $p$ is to balance the large values obtained in the DLD due to the orbits of initial conditions escaping to infinity very fast in this case. We can see in Fig. \ref{DLD_fixIt}A that for $N = 5$ iterations, the method reveals some of the features of the chaotic saddle. However, when we increase the iterations to $N = 10$, see Fig. \ref{DLD_fixIt}B, most of the orbits have escaped to infinity, yielding NaN values when computing DLDs and obscuring completely the phase space structure of the map.

\begin{figure}[htbp]
	\begin{center}
		A)\includegraphics[scale=0.3]{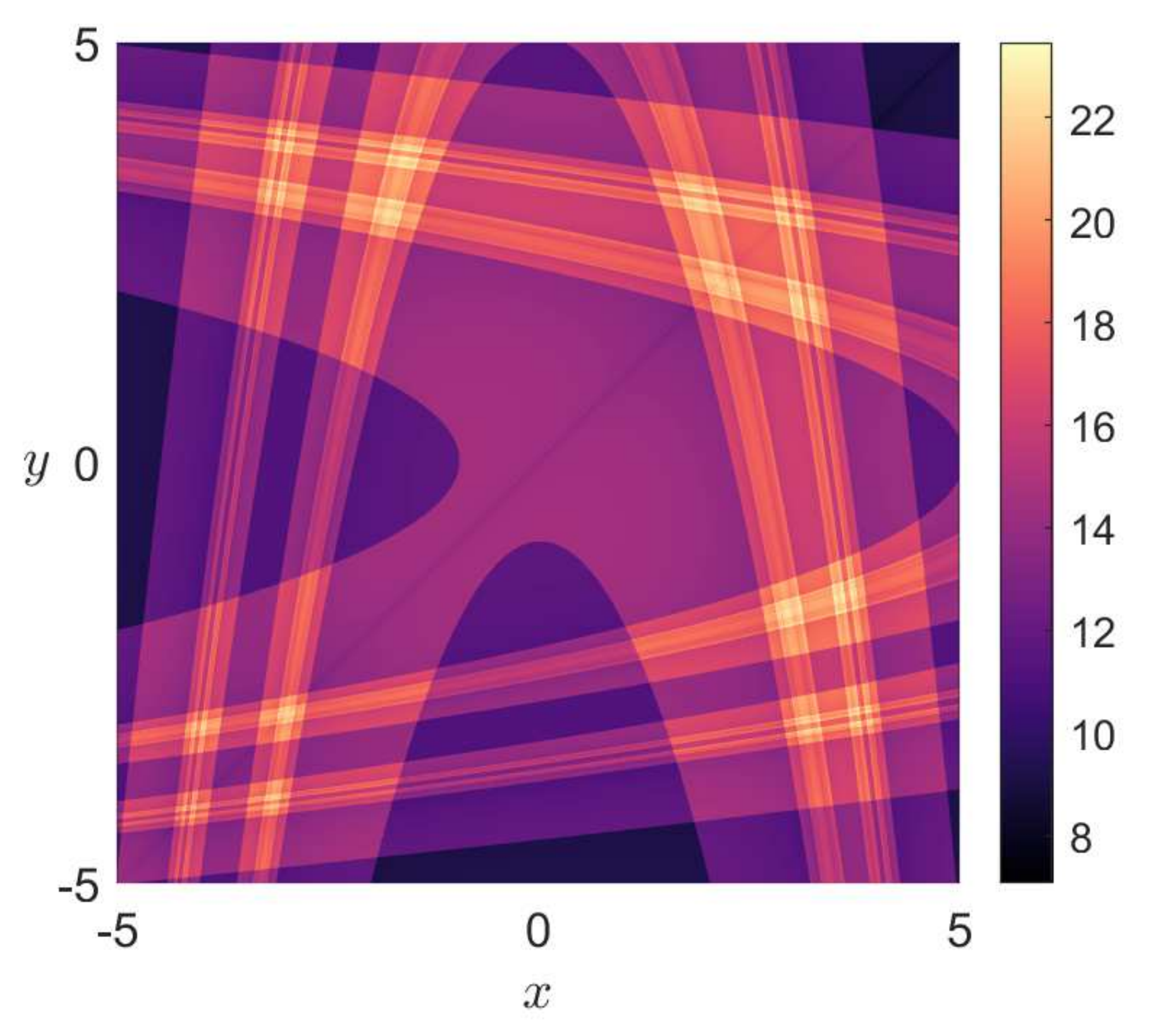}
		B)\includegraphics[scale=0.3]{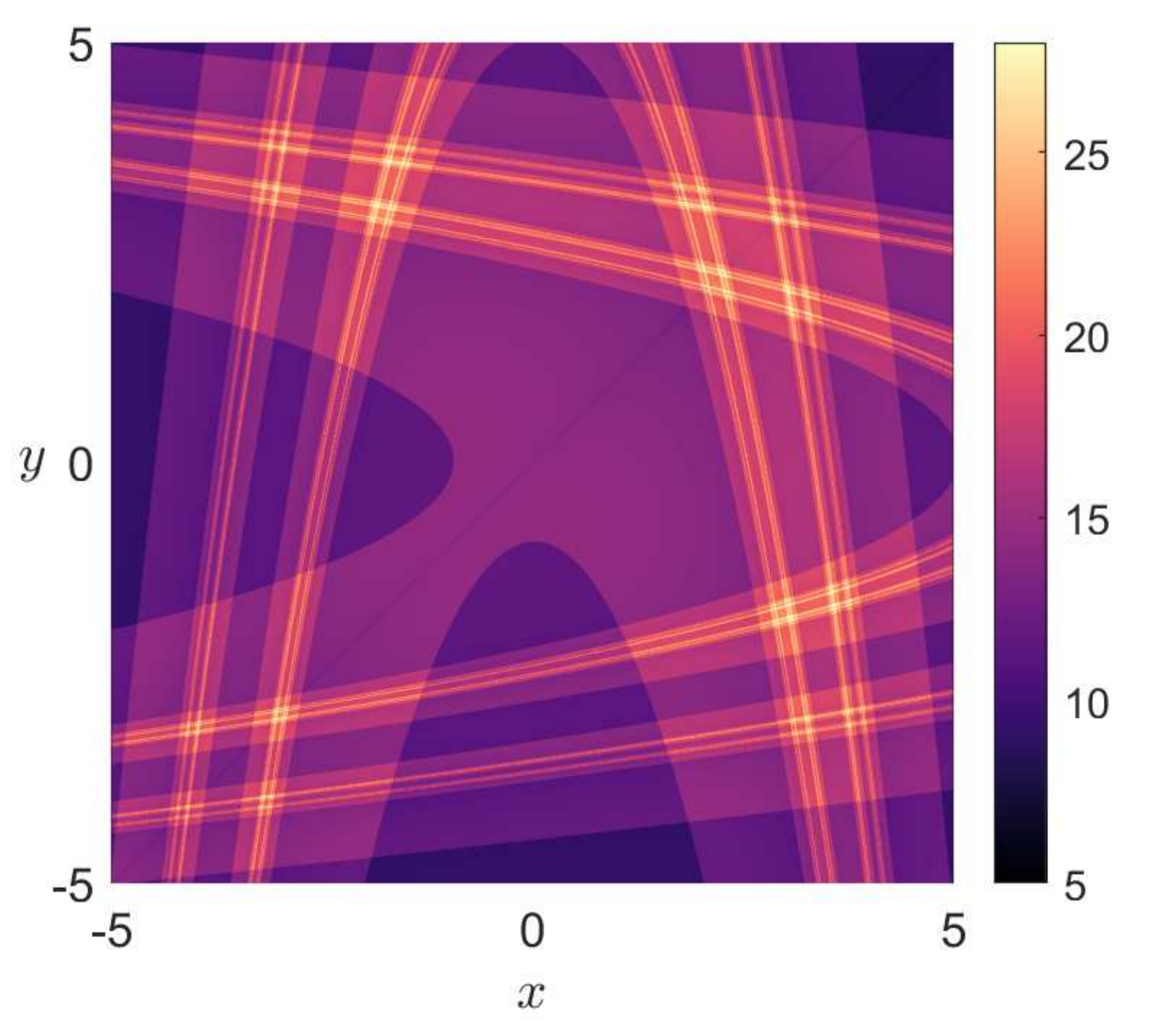}
		C)\includegraphics[scale=0.3]{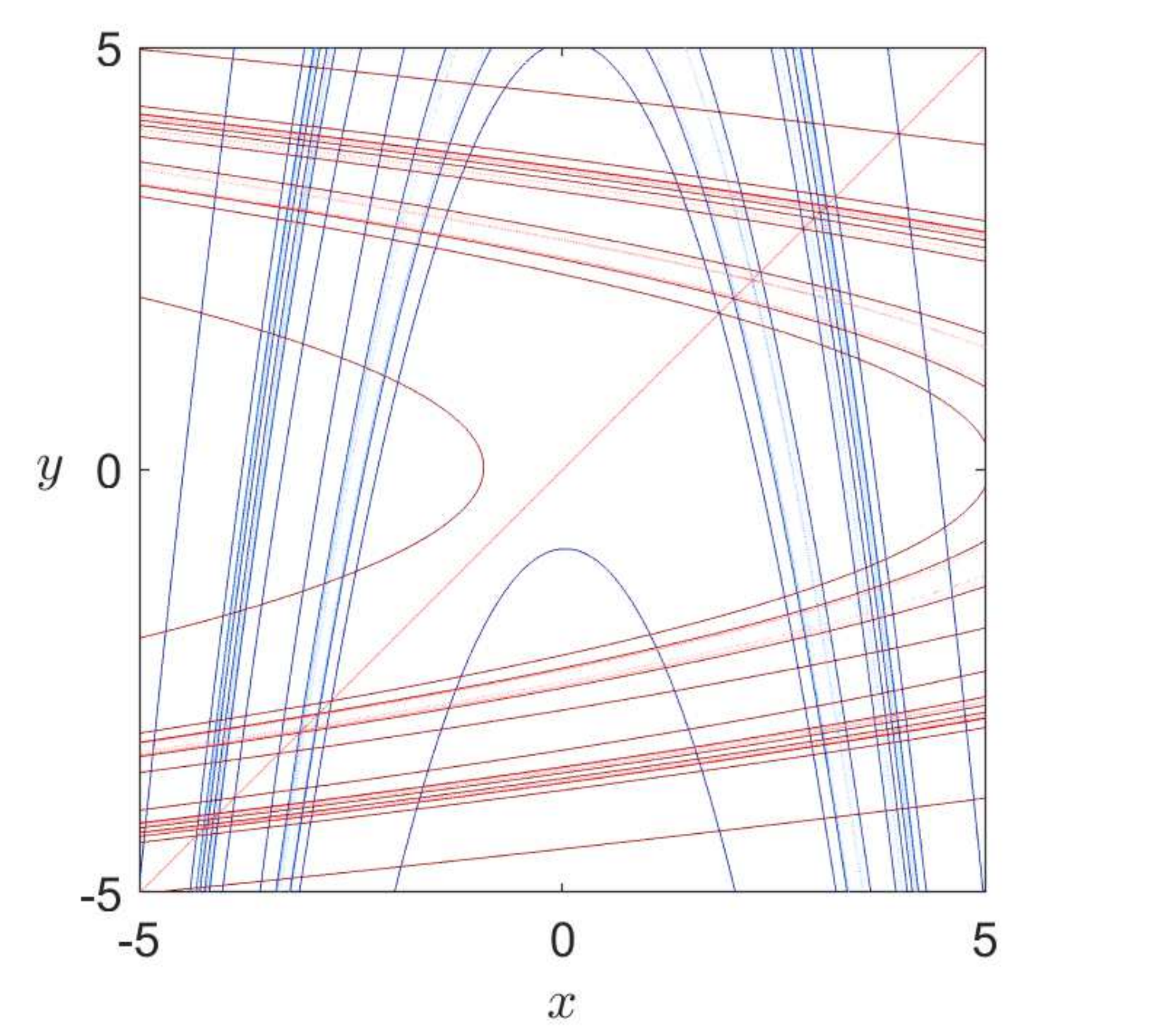}
		D)\includegraphics[scale=0.3]{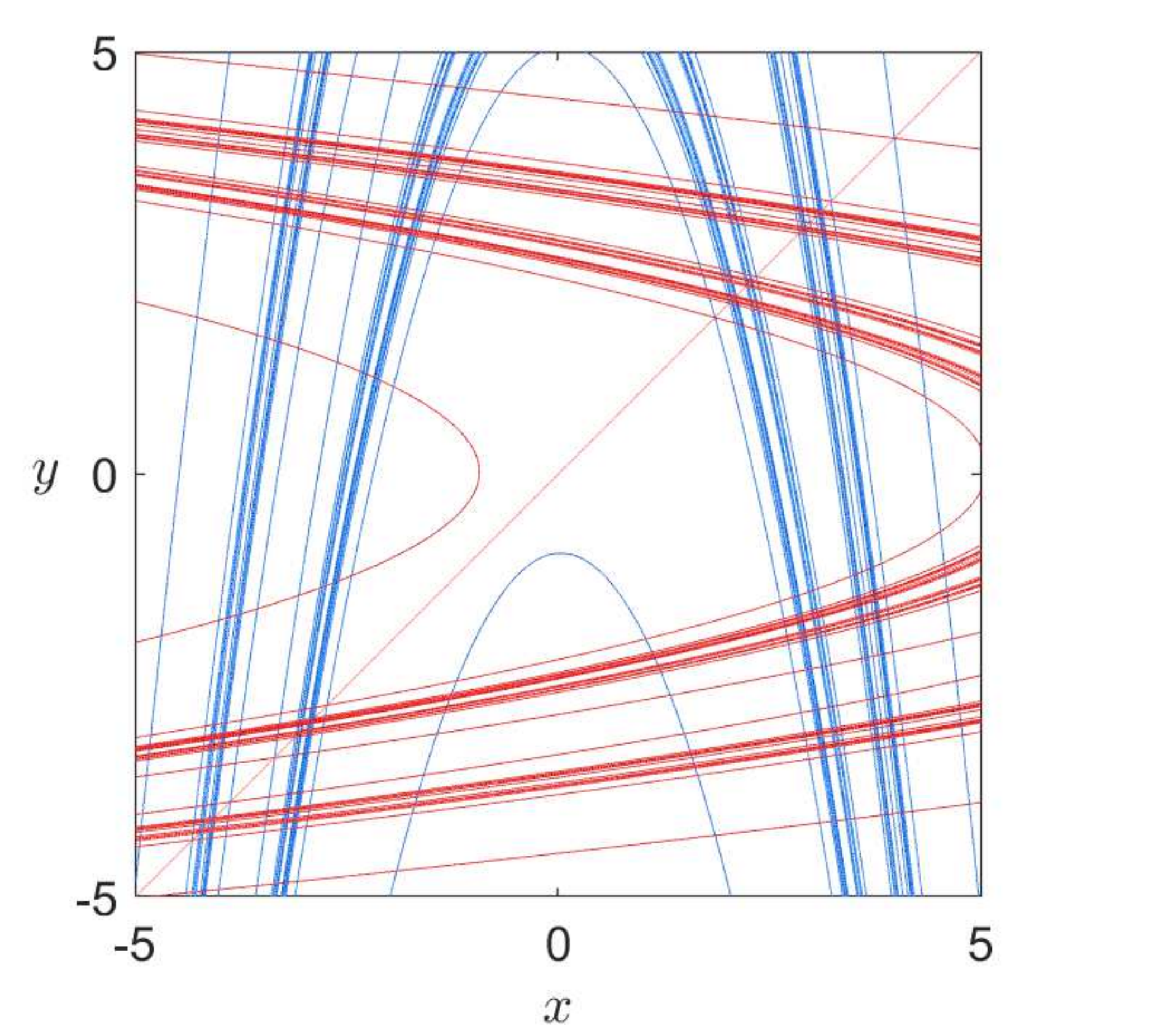}
		E)\includegraphics[scale=0.3]{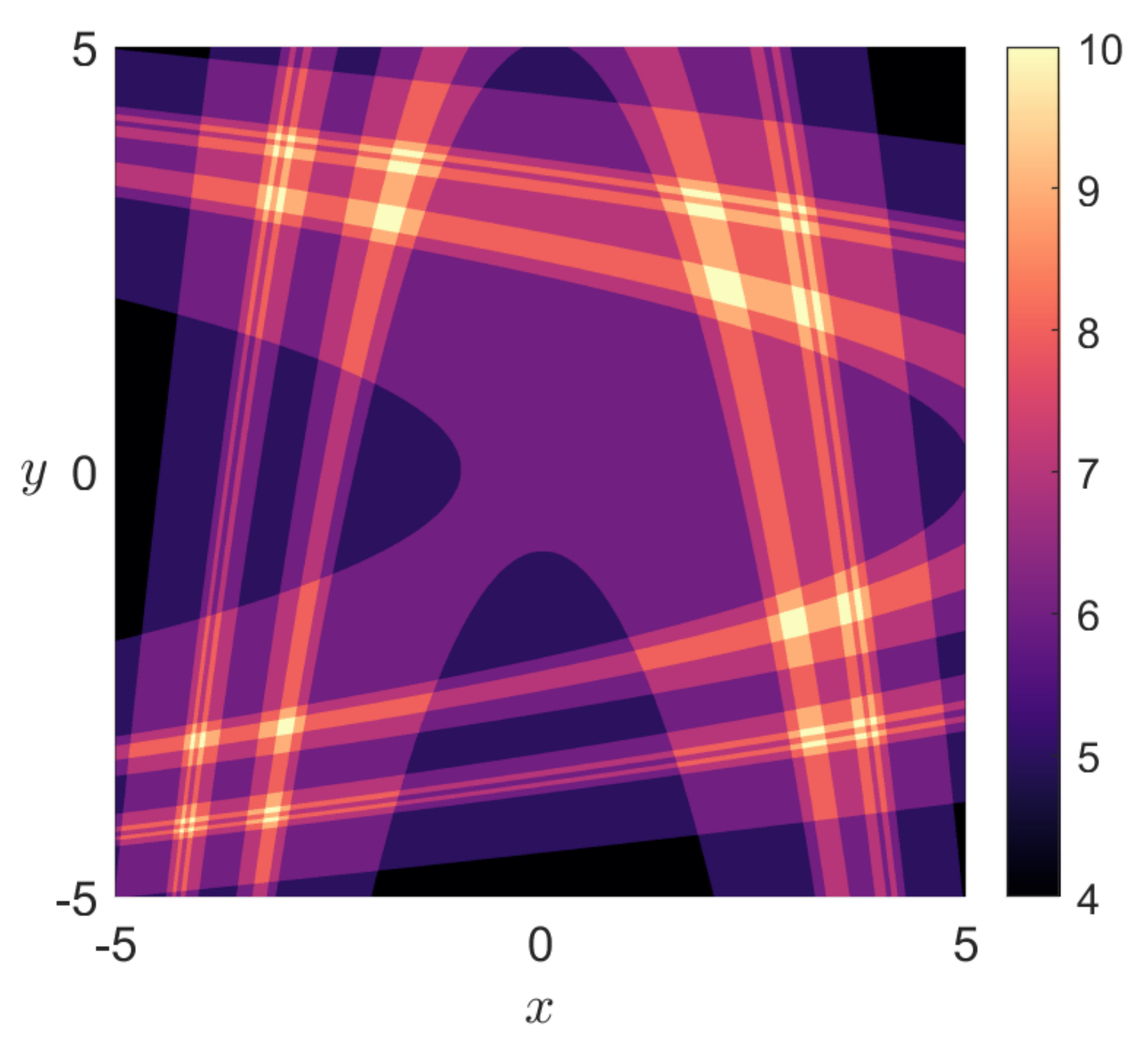}	
		F)\includegraphics[scale=0.3]{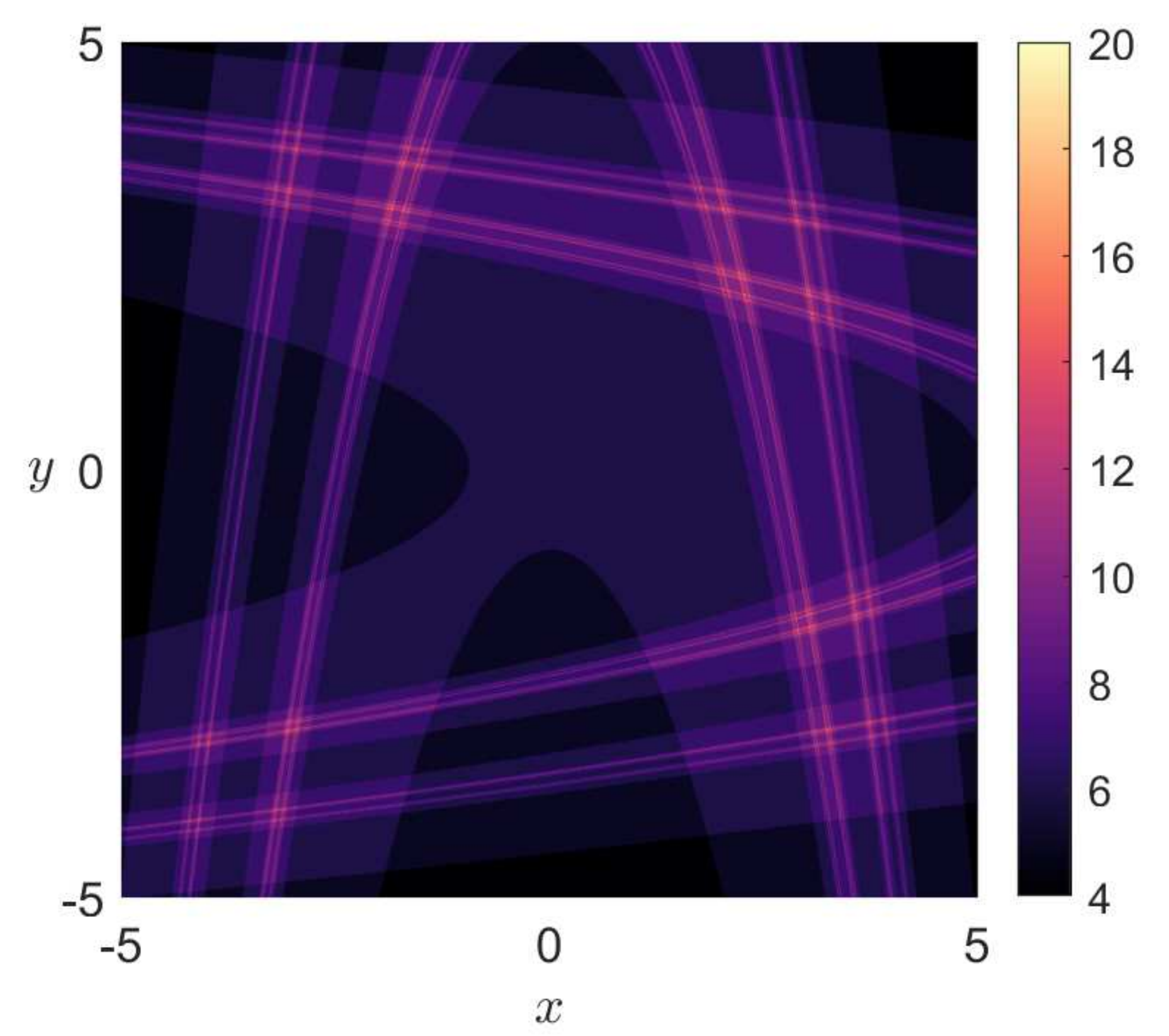}	
	\end{center}
	\caption{Phase space structures of the H\'{e}non map for the model parameters $A = 9.5$ and $B = -1$. The left column corrsponds to $N = 5$ iterations, and the right column is for $N = 10$ iterations. A) and B) VIN-DLD calculated with $p = 0.05$. C) and D) Stable (blue) and unstable (red) manifolds extracted from the gradient of DLD. E) and F) Transit time distributions.}
	\label{varIt_DLD_case1}
\end{figure}

With this example, it becomes clear that stopping the orbits of initial conditions at some point along their evolution might help circumvent this issue. To do so, we fix a circular region of radius $r = 100$ about the origin in the phase plane and apply the variable-iteration definition in Eq. \eqref{VT_DLD}. Therefore, given a fixed number of iterations $N$, we will calculate DLDs along the orbit of an initial condition until we reach the maximum number of iterations $N$, or the trajectory leaves the circular region. We can clearly see in Fig. \ref{varIt_DLD_case1}A that the VIN-DLD, using $p = 0.05$ recovers for $N = 5$, the basic characteristics of the chaotic saddle as the FIN-DLD does. Furthermore, if we increase the number of iterations to $N = 10$ we observe, see Fig. \ref{varIt_DLD_case1}B, that this time the technique succeeds in recovering more features of the intricate and fractal structure of the chaotic saddle. This is expected from the method as the number of iterations gets large, since we are incorporating more information about the future and past history of orbits into the DLD computation and, consequently, this results in a detailed detection of the stable and unstable manifolds. This exapmple demonstrates that the variable iteration number definition of DLDs overcomes the issue encountered by the FIN-DLD when applied to unbounded maps with trajectories escaping to infinity at an increaseing rate. It is important to highlight that, once $\mathcal{D}_p$ has been calculated, we can directly extract from it the stable and unstable manifolds of the map by means of processing the values of its gradient, since these invariant manifolds are mathematically connected with the points at which the DLD scalar field is non-differentiable, as was proved in \citet{carlos}. We illustrate this capability of the method in Figs. \ref{varIt_DLD_case1}C-D. As a validation, we compare the output of $\mathcal{D}_p$ with exit time distributions \cite{meiss1997}, i.e. the total number of iterations (forwards plus backwards) that the orbit of each initial condition remains inside a fixed phase space region. Therefore, the transit time distribution is given by:
\begin{equation}
\mathcal{T}_{\mathbf{x}_0} = N^{+}_{\mathbf{x}_0} + N^{-}_{\mathbf{x}_0}
\end{equation}
where $N^{\pm}_{\mathbf{x}_0}$ are defined in Eq. \eqref{var_its}. Notice that, in order to compute $\mathcal{D}_p$ we need to keep track of the number of iterations for which  the orbit of each initial condition remains inside a given phase space region (before escaping), thus, in the process of calculating VIN-DLD we also obtain simultaneously the transit time distribution. The comparison between both methods shows an excellent agreement as displayed in Figs. \ref{varIt_DLD_case1}E-F.

\begin{figure}[htbp]
	\begin{center}
		A)\includegraphics[scale=0.3]{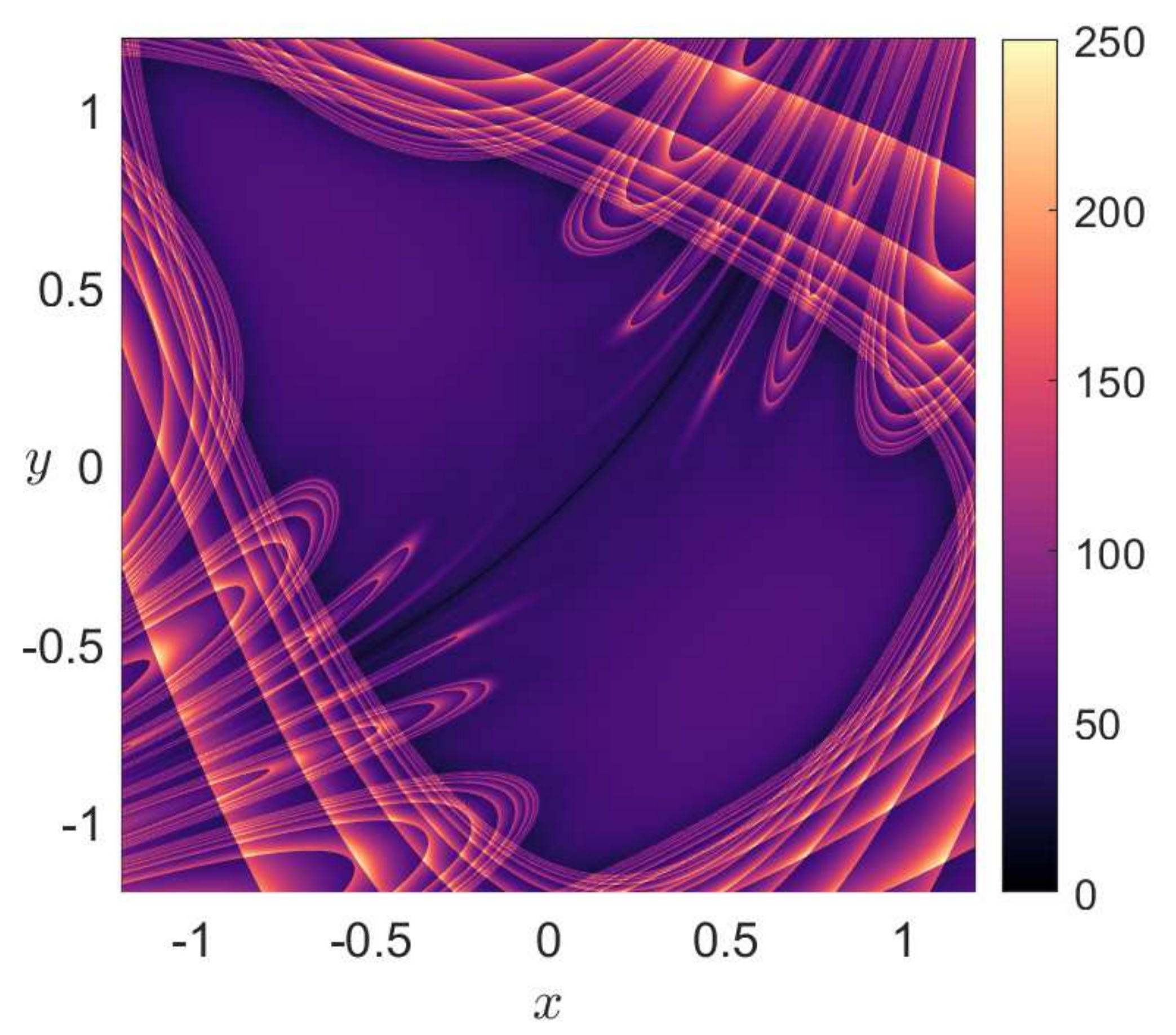}
		B)\includegraphics[scale=0.3]{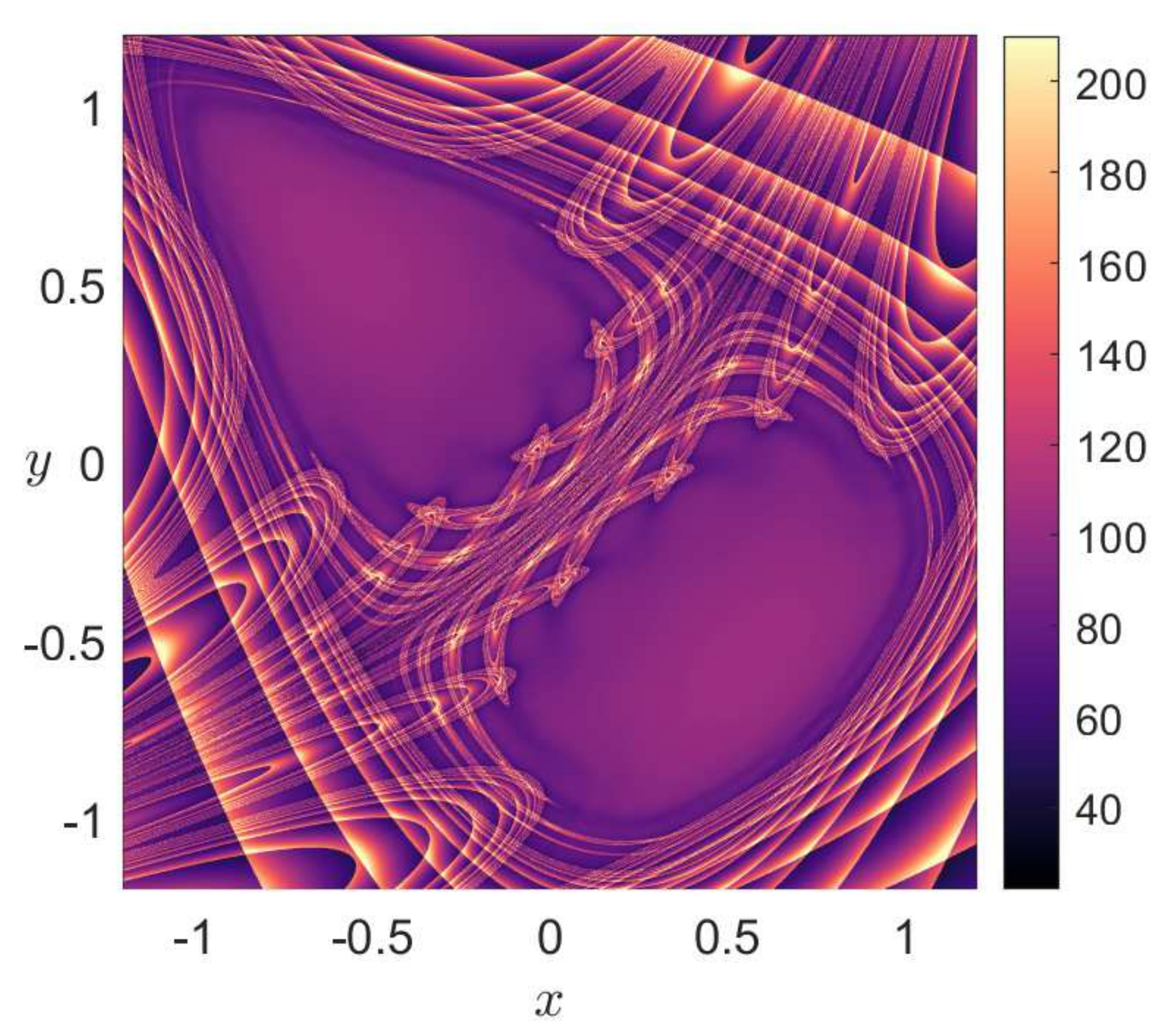}
		C)\includegraphics[scale=0.31]{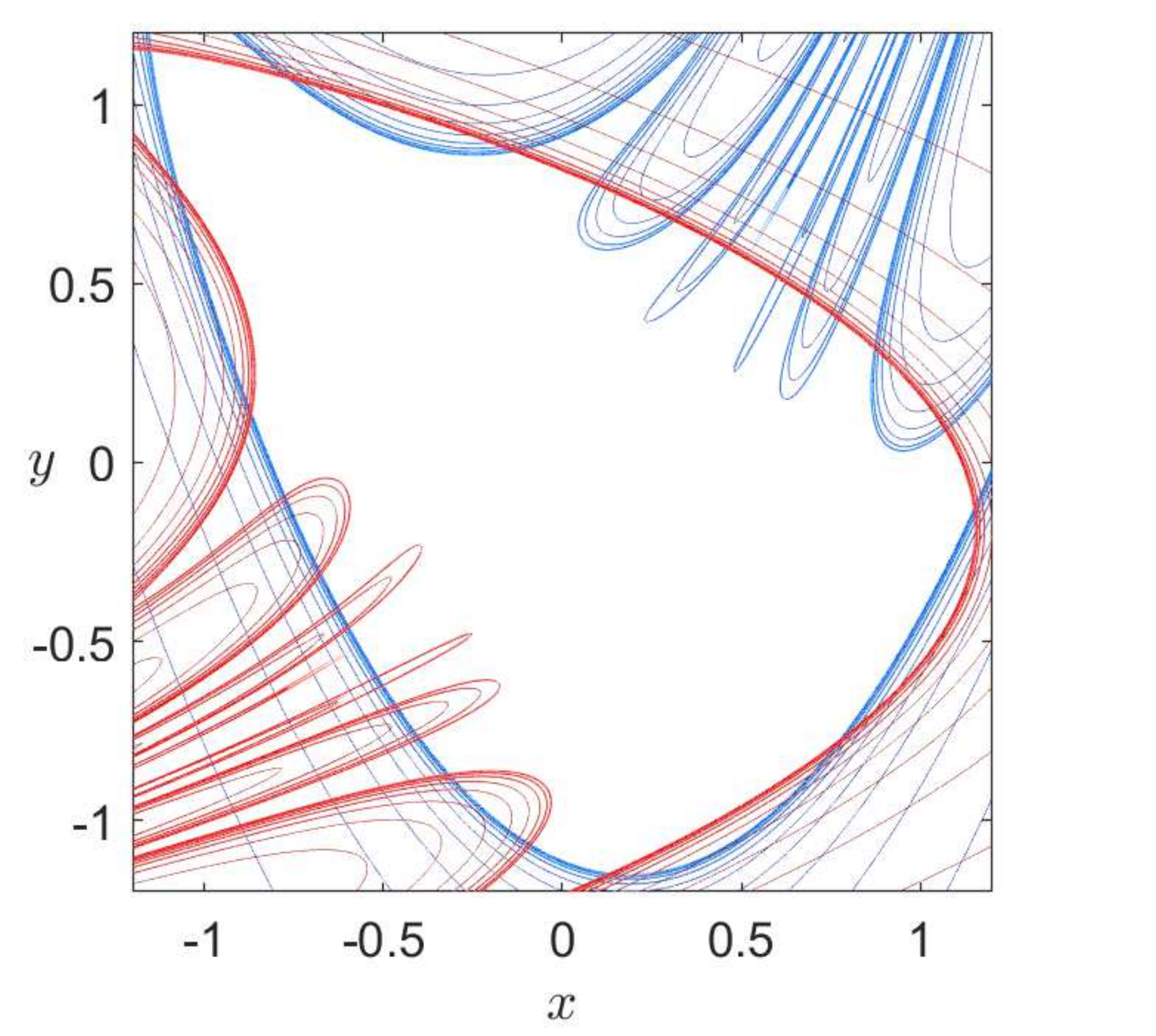}
		D)\includegraphics[scale=0.31]{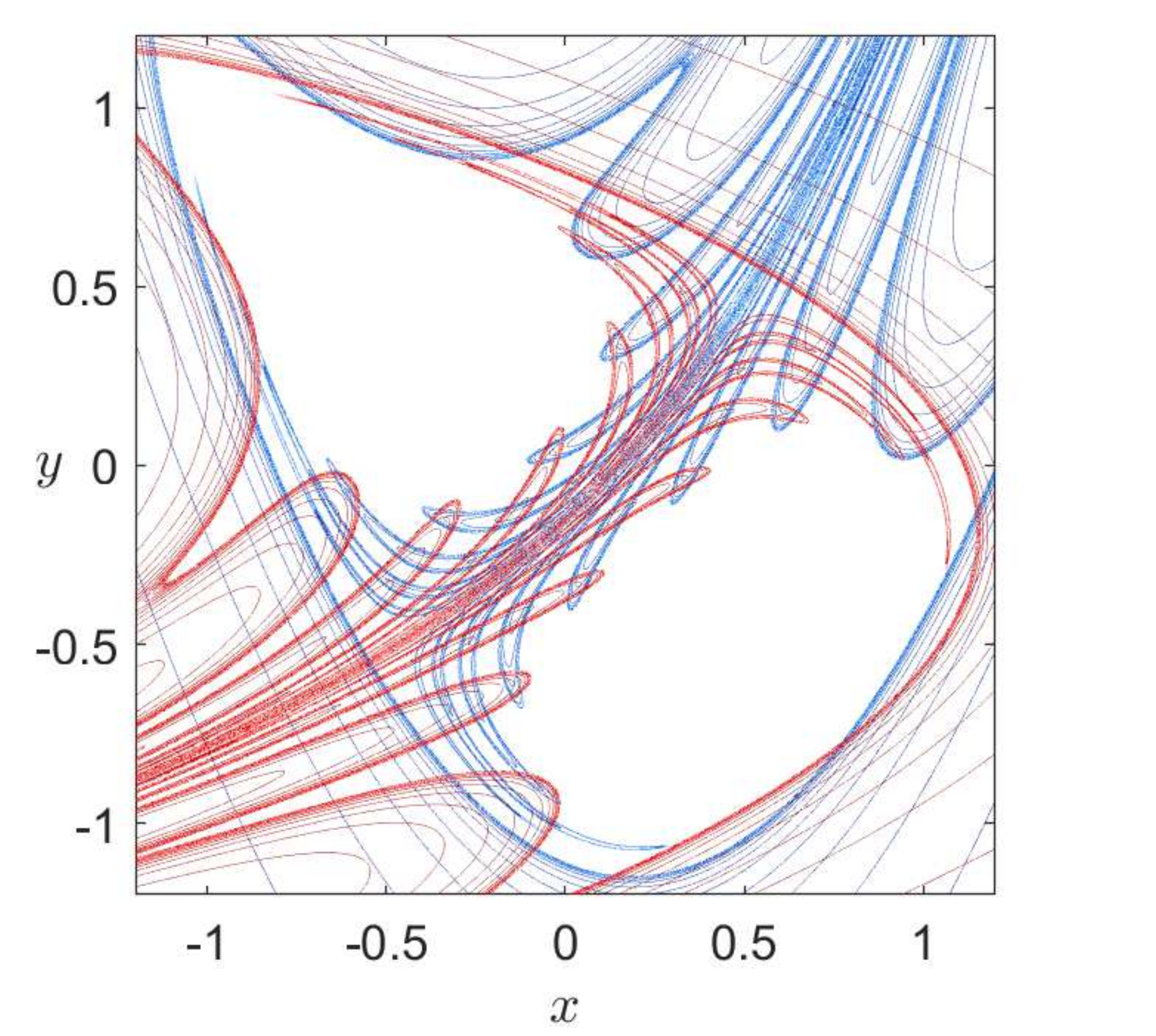}
		E)\includegraphics[scale=0.3]{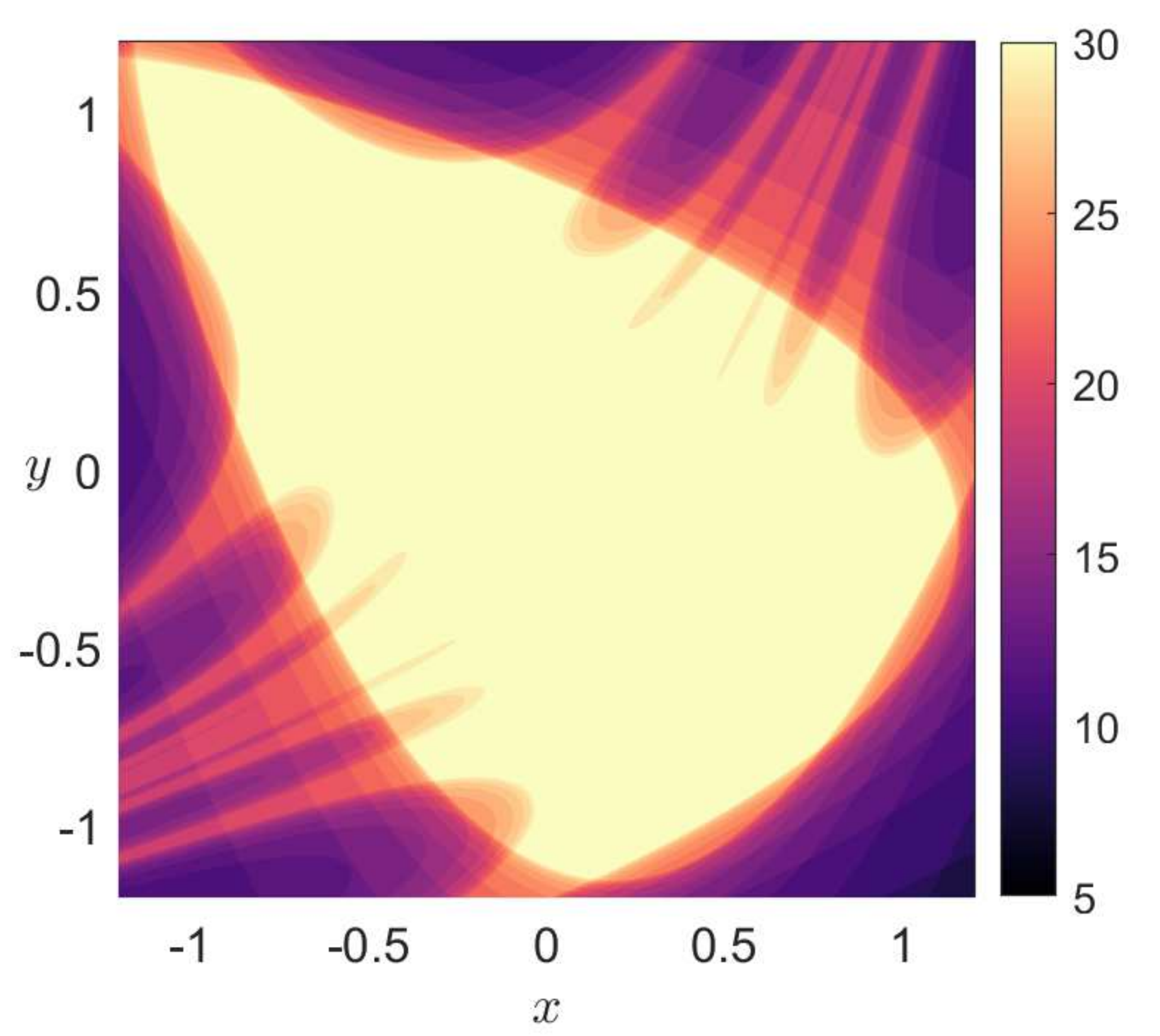}	
		F)\includegraphics[scale=0.3]{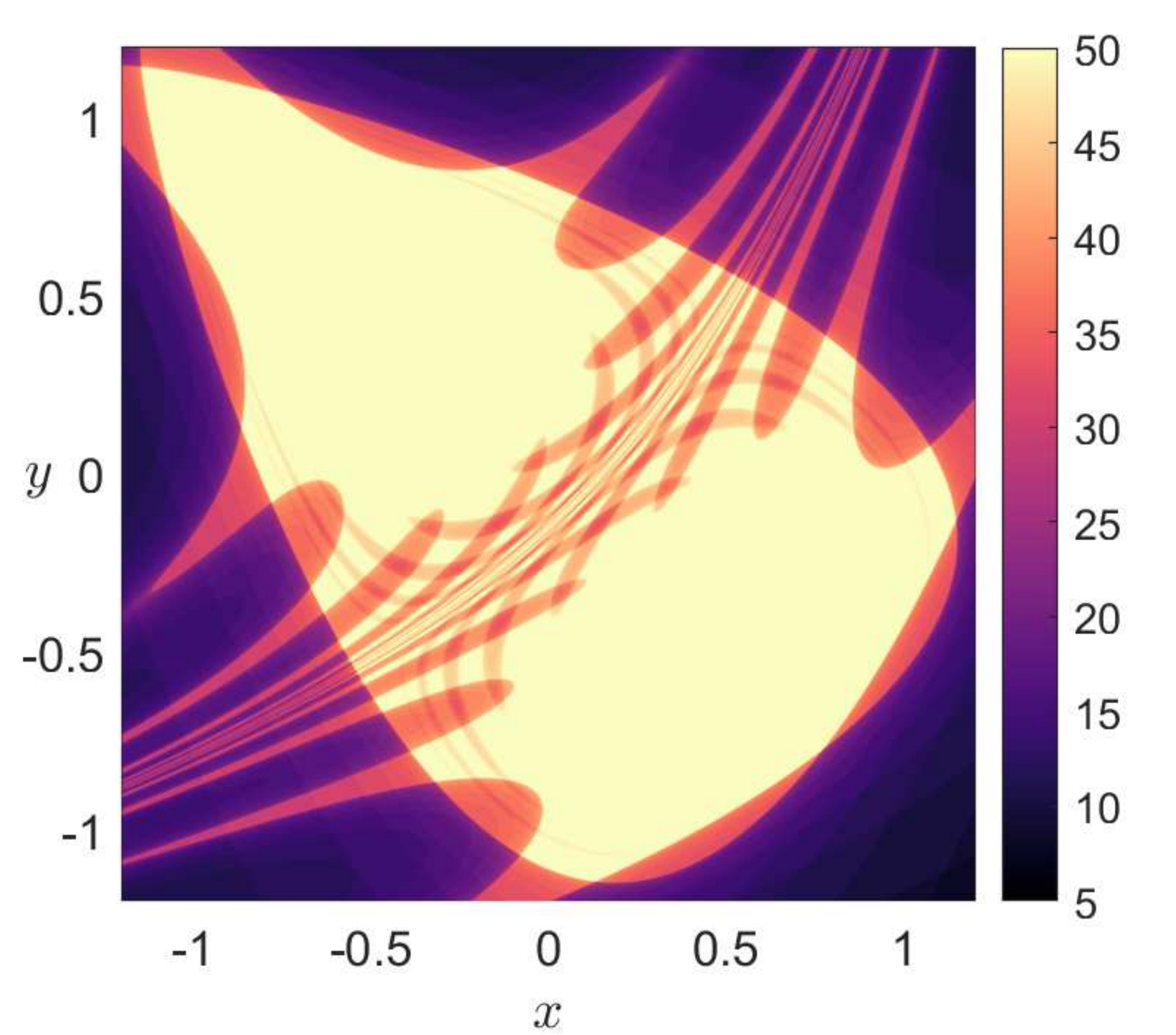}	
	\end{center}
	\caption{Phase space structures of the H\'{e}non map for the model parameters $A = 0.298$ and $B = 1$. The left column corrsponds to $N = 15$ iterations, and the right column is for $N = 25$ iterations. A) and B) VIN-DLD calculated with $p = 0.5$. C) and D) Stable (blue) and unstable (red) manifolds extracted from the gradient of DLD. E) and F) Transit time distributions.}
	\label{varIt_DLD_case2}
\end{figure}

\begin{figure}[htbp]
	\begin{center}
		A)\includegraphics[scale=0.27]{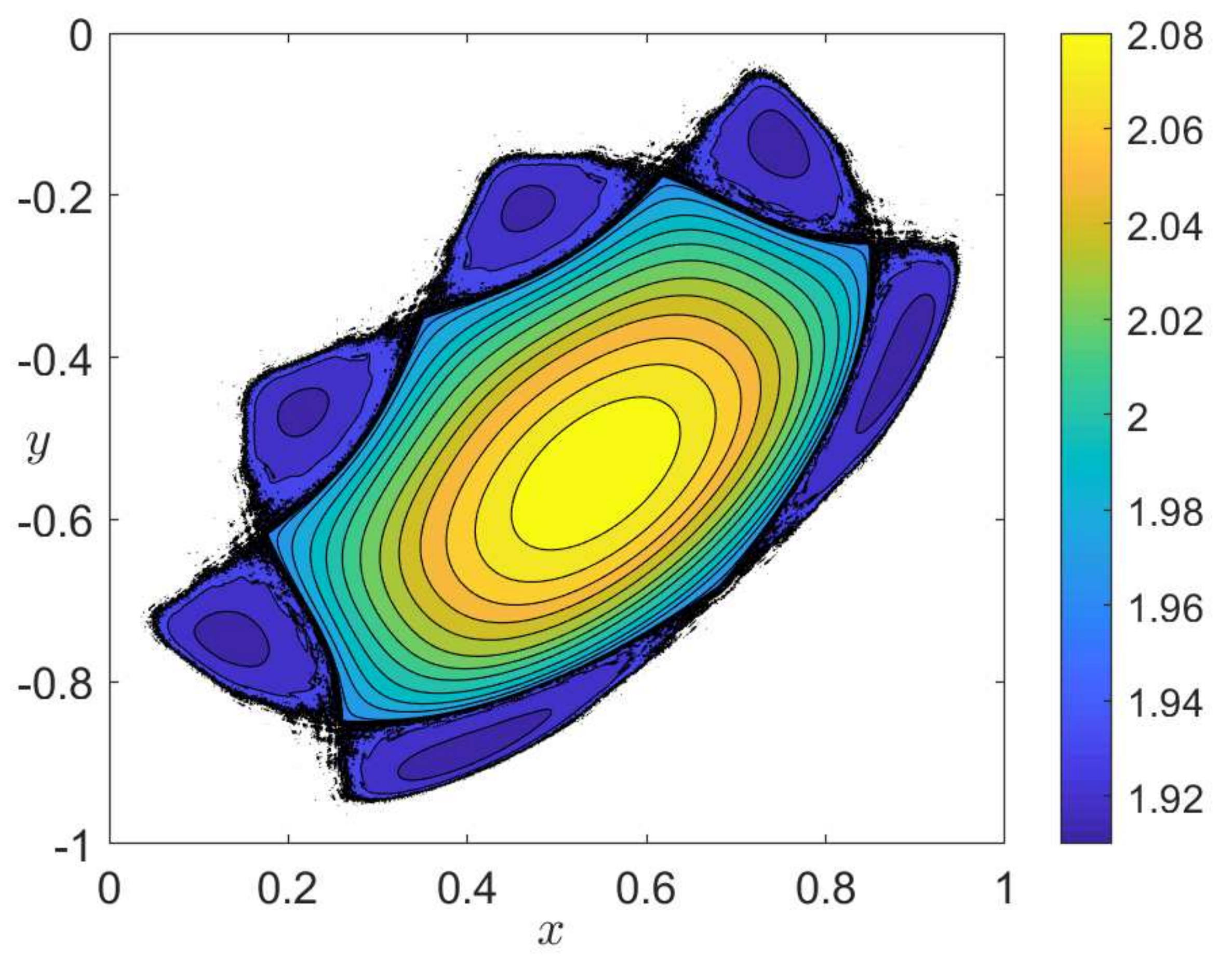}
		B)\includegraphics[scale=0.27]{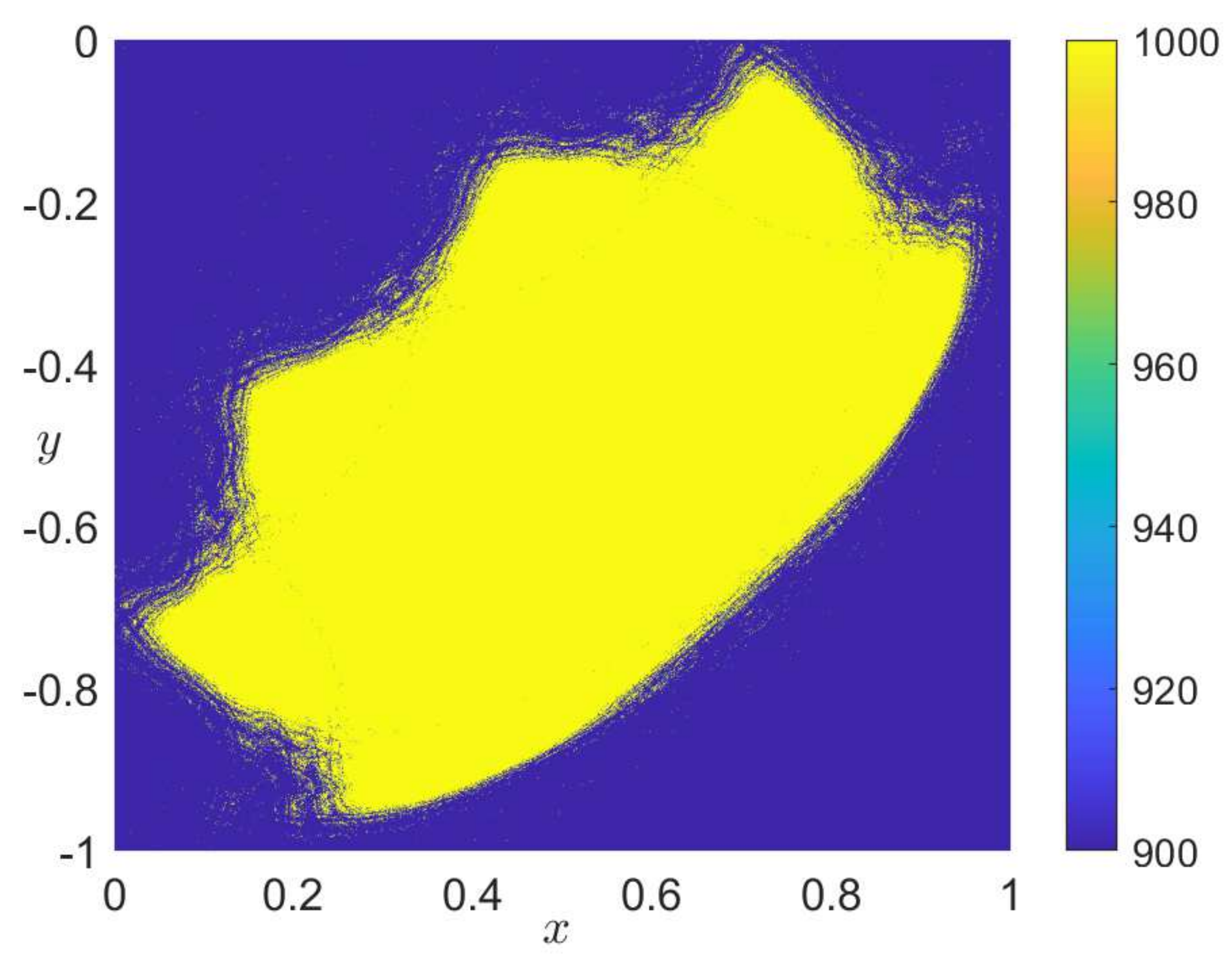}
	\end{center}
	\caption{Phase space KAM tori of the H\'{e}non map for the model parameters $A = 0.298$ and $B = 1$. A) Contour levels of the VIN-DLD calculated with $p = 0.5$ for $N = 500$ iterations. B) Transit time distribution.}
	\label{varIt_DLD_KAM}
\end{figure}

We turn our attention next to analyze the phase space structures of the H\'{e}non map with model parameters $A = 0.298$ and $B = 1$. In this case we know that there is dynamical trapping due to the stickiness of the KAM tori present in two regularity islands of the phase space, wich are surrounded by the homoclinic and heteroclinic intersections of the stable and unstable manifolds of periodic saddles in the chaotic sea \cite{oliveira2019}. In order to uncover the invariant manifolds we apply $\mathcal{D}_p$ with $p = 0.5$ using $N = 15$ and $N = 25$ iteartions. We illustrate in this way how the method captures more features of the stable and unstable manifolds as the number of iterations is increased. The result of this calculation is displayed in Figs. \ref{varIt_DLD_case2}A-B. As we have explained, the stable and unstable manifolds are revealed at points of the scalar field, known in the literature as singular features, where $D_{p}$ is non-differentiable (the gradient becomes unbounded). We also demonstrate the potential of extracting the invariant stable and unstable manifolds of the map from the gradient of $\mathcal{D}_p$ in Figs. \ref{varIt_DLD_case2}C-D. Moreover, observe that the islands of regularity corresponding to KAM tori are recovered by regions where the $\mathcal{D}_p$ function is smooth. In fact, a mathematical connection has been established between time-averaged Lagrangian descriptors \cite{lopesino2017} and KAM tori through Birkhoff's Ergodic Partition Theorem  \cite{mezic1999} under certain assumptions. Despite the fact that this proof has only been given for continuous-time dynamical systems, since Eq. \eqref{DLD} can be interpreted as a discretized version of the $p$-norm definition of LDs inntroduced in \cite{lopesino2017}, this property will also hold for maps. Therefore, if we define the iteration-averaged DLD as,
\begin{equation}
\langle \mathcal{D}_{p} \rangle = \dfrac{1}{N} \mathcal{D}_{p}\left(\mathbf{x}_0,N\right)
\end{equation}
then the contours of $\langle \mathcal{D}_{p} \rangle$ when convergence has been reached as the number of iterations $N \to \infty$ correspond to invariant sets in the phase space. With this approach, we can use $\mathcal{D}_p$ to easily recover the KAM tori as we demonstrate in Fig. \ref{varIt_DLD_KAM}A, where we have represented the contours of $\langle \mathcal{D}_{p} \rangle$ calculated for $N = 500$ iterations. We compare this output to that obtained by plotting the transit time distribution, see Fig. \ref{varIt_DLD_KAM}B. Since orbits are trapped in the tori forever, the exit-time plot is flat all over the regularity island and thus transit times do not give further information about the structure of the KAM tori in that region. To summarize, we have illustrated that VIN-DLD has some advantages with respect to the analysis of transit time distributions. First, it is straightforward to extract the stable and unstable manifolds of the map directly from the computation of $||\nabla \mathcal{D}_p||$, and second, it simultaneously depicts the KAM tori associated to islands of regularity when the number of iterations is chosen large enough, thanks to its connection with Birkhoff's Ergodic Partition Theorem.

\begin{figure}[htbp]
	\begin{center}
		A)\includegraphics[scale=0.32]{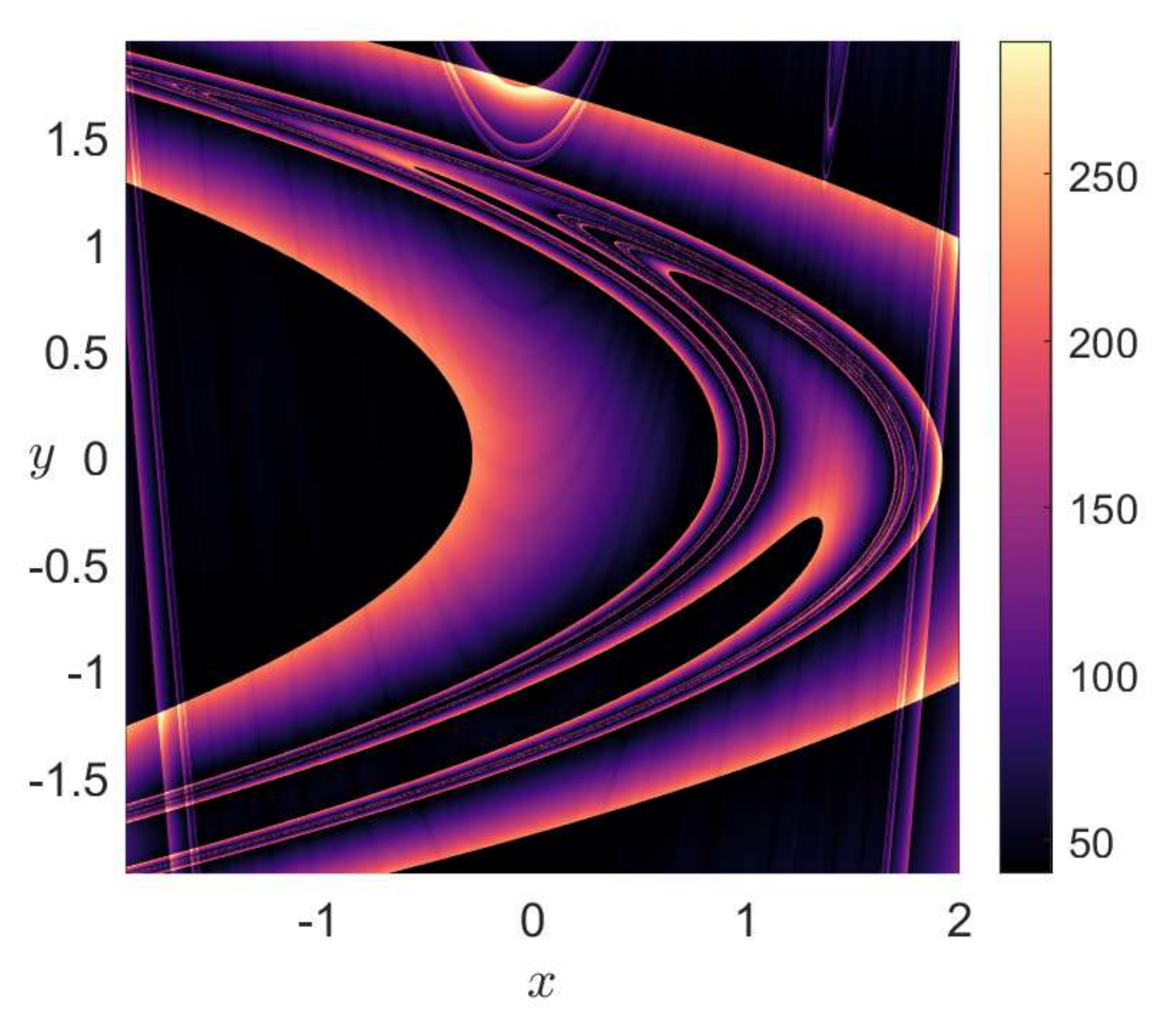}
		B)\includegraphics[scale=0.32]{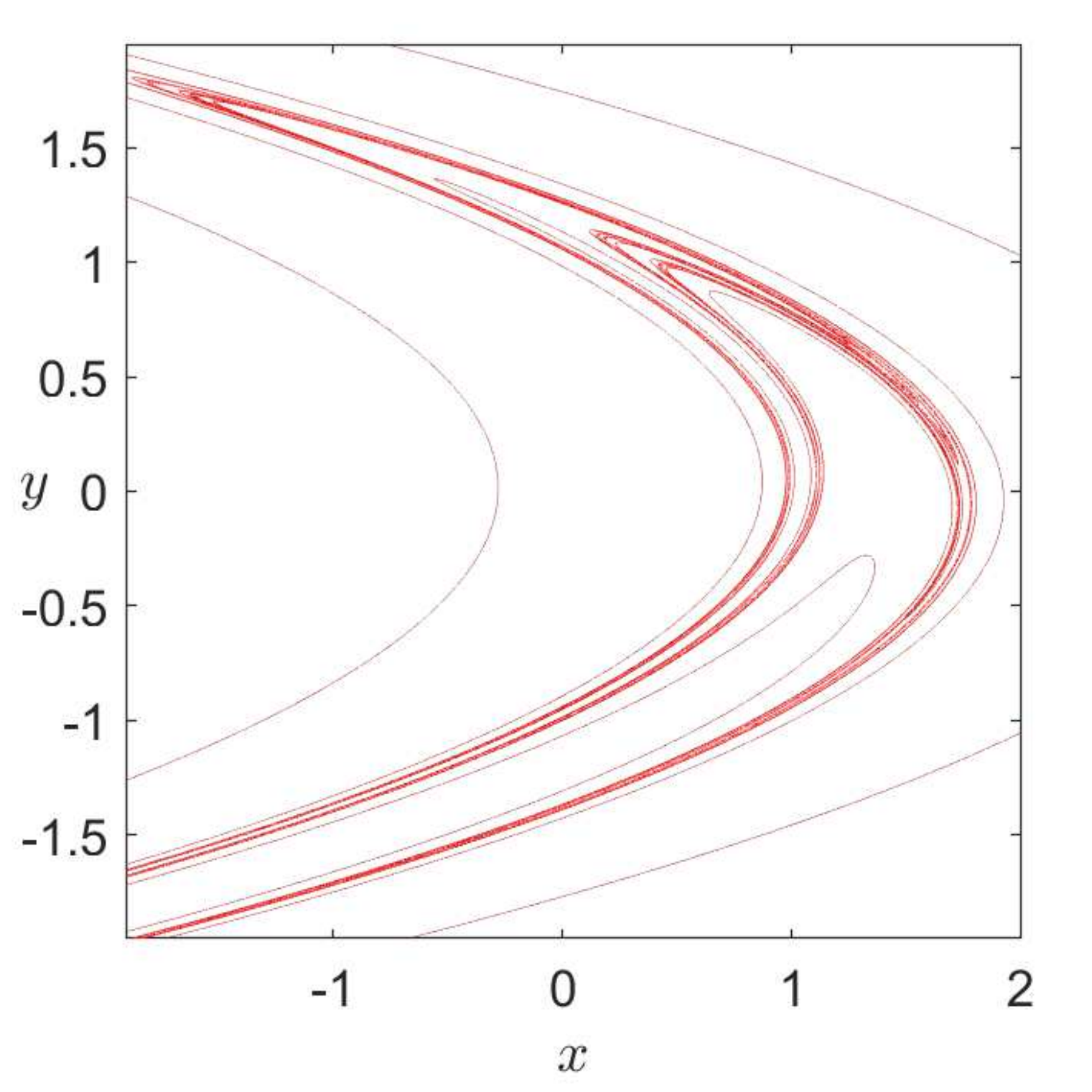}	
		C)\includegraphics[scale=0.32]{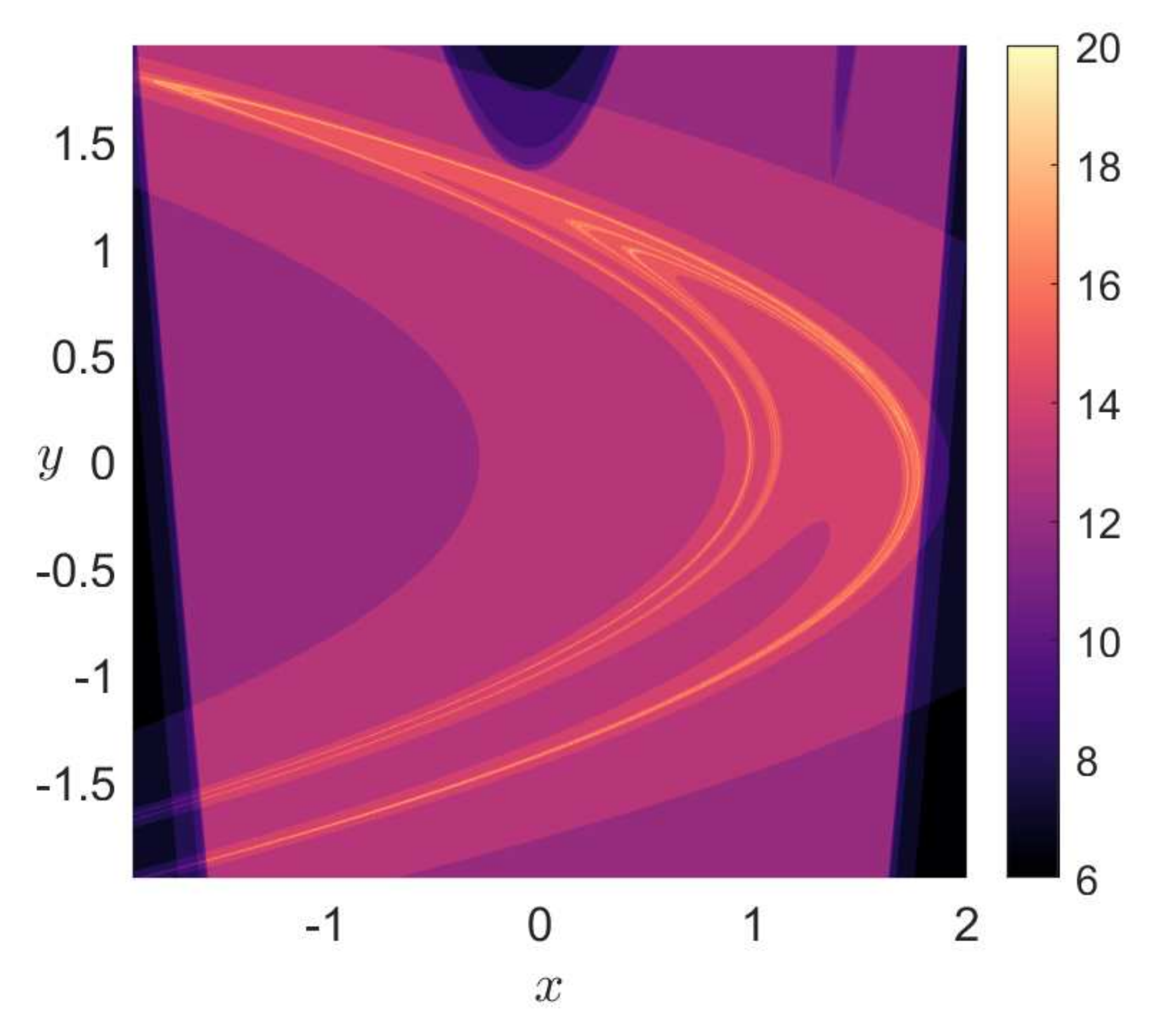}
	\end{center}
	\caption{Strange attractor of the H\'{e}non map for the model parameters $A = 1.4$ and $B = 0.3$. A) Variable iteration number DLD calculated with $p = 0.5$ for $N = 10$ iterations. B) Unstable manifold extracted from the gradient of $\mathcal{D}_p$. C) Transit time distribution.}
	\label{varIt_DLD_strAtt}
\end{figure}

To finish, we will apply $\mathcal{D}_p$ with $p = 0.5$ for $N=10$ iterations to uncover the phase space structure of the H\'{e}non map with model parameters $A = 1.4$ and $B = 0.3$. For this case, we know that the map does not preserve area, and that it has a strange attractor that has been widely studied in the literature \cite{henon1976}. In Fig. \ref{varIt_DLD_strAtt} we display the results of this computation, confirming that the method successfully unveils the intricate structure of the attractor. Moreover, the structure of the strange attractor can be recovered completely in a simple way from the gradient of the backward iteration of the method which, as we have explained, captures the unstable manifolds of the map.

\section{Conclusions}
\label{sec:conc}

In this work we have extended the method of Discrete Lagrangian Descriptors with the purpose of improving its capability for revealing the phase space structure of unbounded maps, where trajectories can escape to infinity at an increasing rate. In order to achieve this goal, and given a fixed number of maximum iterations, we have modified the definition of DLDs so that its value is accumulated along the orbit of each initial condition up to the fixed number of iterations, or until the trajectory leaves a certain region in the plane. In order to illustrate the success of this alternative variable iteration number definition to unveil the template of geometrical phase space structures we have applied it to the well-known H\'{e}non map, for which the original FIN-DLD has issues in order to fully reveal the intricate geometry of the stable and unstable manifolds, and also the presence of a chaotic saddle and a strange attractor for some values of the model parameters. Finally, and as a validation, we have compared our simulations of DLDs with the classical escape time plots used in the Dynamical System's literature to uncover invariant manifolds. Our results are in excellent agreement with the well-studied properties of the H\'{e}non map and therefore we are confident that this technique will surely provide the Dynamical System's community with a valuable resource to explore the dynamics and dynamical characteristics of discrete time systems.

\section*{Acknowledgments}

I would like to thank Prof. Stephen Wiggins for his useful comments and suggestions that helped me improve the contents of this paper. The author also acknowledges the financial support received from the EPSRC Grant No. EP/P021123/1 and the Office of Naval Research Grant No. N00014-01-1-0769 for his research visits over the past year at the School of Mathematics, University of Bristol. This work is the result of many fruitful discussions with Prof. Stephen Wiggins during that period.

\bibliographystyle{natbib}
\bibliography{SNreac}

\end{document}